\documentclass{www2013-accepted}

\usepackage{amsmath}
\usepackage{amssymb}
\usepackage{graphicx}
\usepackage{array}
\usepackage{color}
\usepackage{times}

\DeclareMathOperator*{\argmin}{\mathrm{argmin}}
\newcommand{\eq}[1]{(eq.~\ref{#1})}
\newcommand{\xxhdr}[1]{\vspace{0mm}\noindent{{\bf #1}}}
\newcommand{\xhdr}[1]{\subsubsection*{#1}}

\newcommand{\modela}{a}
\newcommand{\modelb}{b}
\newcommand{\modelc}{c}
\newcommand{\modeld}{d}

\begin{document}

\title{From Amateurs to Connoisseurs: \\ Modeling the Evolution of User Expertise \\ through Online Reviews}

\numberofauthors{2}
\author{
\alignauthor
Julian McAuley\\
       \affaddr{Stanford University
       \email{jmcauley@cs.stanford.edu}}
\alignauthor
Jure Leskovec\\
       \affaddr{Stanford University
       \email{jure@cs.stanford.edu}}
}

\maketitle

\begin{abstract}
Recommending products to consumers means not only understanding their \emph{tastes}, but also understanding their level of \emph{experience}. For example, it would be a mistake to recommend the iconic film \emph{Seven Samurai} simply because a user enjoys other action movies; rather, we might conclude that they will \emph{eventually} enjoy it---once they are ready. The same is true for beers, wines, gourmet foods---or any products where users have acquired tastes: the `best' products may not be the most `accessible'. Thus our goal in this paper is to recommend products that a user will enjoy \emph{now}, while acknowledging that their tastes may have changed over time, and may change again in the future. We model how tastes change due to the very act of consuming more products---in other words, as users become more \emph{experienced}. We develop a latent factor recommendation system that explicitly accounts for each user's level of experience. We find that such a model not only leads to better recommendations, but also allows us to study the role of user experience and expertise on a novel dataset of fifteen million beer, wine, food, and movie reviews.
\end{abstract}

\category{H.3.3}{Information Search and Retrieval}{Information Search and Retrieval}
\keywords{recommender systems, expertise, user modeling}

\section{Introduction}
\label{sec:intro}

\begin{quote}
{\em ``Even when experts all agree, they may well be mistaken''}\flushright -- Bertrand Russell
\end{quote}

In order to predict how a user will respond to a product, we must understand the tastes of the user and the properties of the product. We must also understand how these properties change and evolve over time. As an example, consider the \emph{Harry Potter} film series: adults who enjoy the films for their special effects may no longer enjoy them in ten years, once their special effects are obsolete; children who enjoy the films today may simply outgrow them in ten years; \emph{future} children who watch the films in ten years may not enjoy them, once Harry Potter has been supplanted by another wizard.

This example highlights three different mechanisms that cause perceptions of products to change. Firstly, such change may be tied to the age of the \emph{product}. Secondly, it may be tied to the age (or development) of the \emph{user}. Thirdly, it may be tied to the state (or zeitgeist) of the \emph{community} the user belongs to.

These mechanisms motivate different models for temporal dynamics in product recommendation systems. Our goal in this paper is to propose models for such mechanisms, in order to assess which of them best captures the temporal dynamics present in real product rating data.

A variety of existing works have studied the evolution of products and online review communities. The emergence of new products may cause users to change their focus \cite{kolter07}; older movies may be treated more favorably once they are considered `classics' \cite{koren10}; and users may be influenced by general trends in the community, or by members of their social networks \cite{hao11}.

However, few works have studied the \emph{personal development} of users, that is, how users' tastes change and evolve as they gain knowledge, maturity, and experience. A user may have to be exposed to many films before they can fully appreciate (by awarding it a high rating) \emph{Citizen Kane}; a user may not appreciate a \emph{Roman\'ee-Conti} (the `Citizen Kane' of red wine) until they have been exposed to many inferior reds; a user may find a strong cheese, a smokey whiskey, or a bitter ale unpalatable until they have developed a tolerance to such flavors. The very act of consuming products will cause users' tastes to change and evolve. Developing new models that take into account this novel viewpoint of user evolution is one of our main contributions.

We model such `personal development' through the lens of user \emph{experience}, or \emph{expertise}. Starting with a simple definition, experience is some quality that users gain over time, as they consume, rate, and review additional products. The underlying hypothesis that we aim to model is that users with similar levels of experience will rate products in similar ways, even if their ratings are temporally far apart. In other words, each user evolves on their own `personal clock'; this differs from other models of temporal dynamics, which model the evolution of user and product parameters on a single timescale \cite{koren10,xiang,xiong}.

Naturally, some users may already be experienced at the time of their first review, while others may enter many reviews while failing to ever become experienced. By individually learning for each user the rate at which their experience progresses, we are able to account for both types of behavior.

Specifically, we model each user's level of experience using a series of latent parameters that are constrained to be monotonically non-decreasing as a function of time, so that each user becomes more experienced (or stays at least as experienced) as they rate additional products. We learn latent-factor recommender systems for different experience levels, so that users `progress' between recommender systems as they gain experience.

`Experience' and `expertise' are \emph{interpretations} of our model's latent parameters. In our context they simply refer to some unobserved quantity of a user that increases over time as they consume and review more products. Intuitively, our monotonicity requirement constrains users to evolve in the same `direction', so that what we are really learning is some property of user evolution that is common to all users, regardless of when they arrive in the community. We perform extensive qualitative analysis to argue that `experience' is a reasonable interpretation of such parameters. In particular, our goal is not to say whether experienced/expert users are `better' or `more accurate' at rating products. We simply model the fact that users with the same level of experience/expertise rate products in a \emph{similar} way.

Our experimental findings reveal that modeling the personal evolution of each user is highly fruitful, and often beats alternatives that model evolution at the level of products and communities. Our models of user experience also allow us to study the differences between experienced and novice users, and to discover \emph{acquired tastes} in novel corpora of fifteen million beer, wine, food, and movie reviews.

\subsection*{A Motivating Example}
We demonstrate the evolution of user tastes and differences between novices and experts (i.e., experienced users) by considering the beer-rating website \emph{RateBeer}, which consists of around three million beer reviews.

In this data we find a classic example of an acquired taste: hops. The highest-rated beers on the website are typically the hoppiest (American Pale Ales, India Pale Ales, etc.); however, due to their bitterness, such beers may be unpalatable to inexperienced users. Thus we might argue that even if such beers are the \emph{best} (i.e., the highest rated), they will only be \emph{recognized} as the best by the most experienced members of the community.

\begin{figure}
\begin{center}
 \includegraphics[width=\columnwidth]{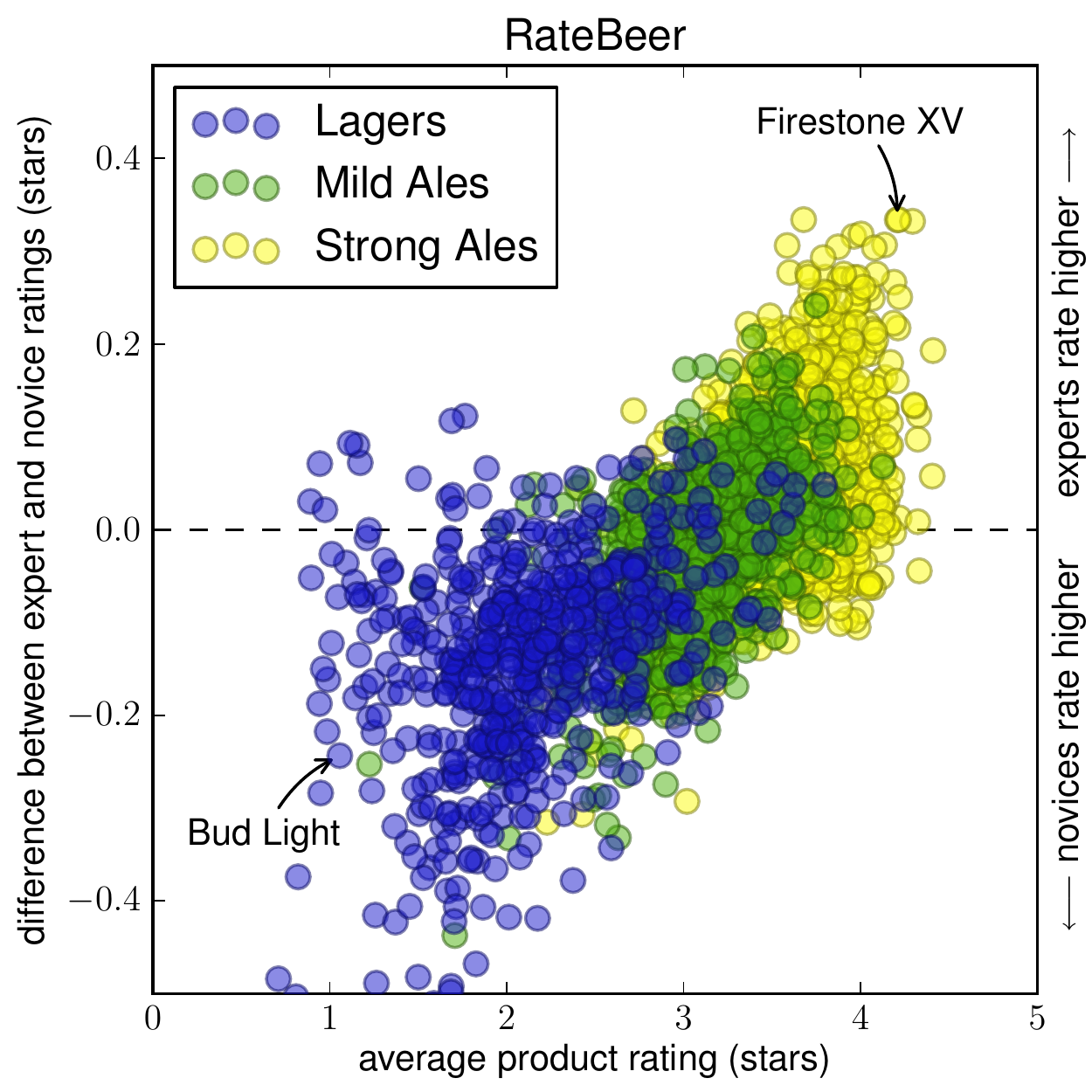}
\end{center}
\caption{
Beloved products are most beloved by experts; hated products are most hated by experts.
\label{fig:genreVexp}
}
\end{figure}

Figure \ref{fig:genreVexp} examines the relationship between product ratings, user experience level, and hoppiness, on \emph{RateBeer} data. Beers of three types are considered (in increasing order of hoppiness): lagers, mild ales, and strong ales. The x-axis shows the average rating of products on the site (out of 5 stars), while the y-axis shows the \emph{difference} between expert and novice ratings. `Experts' are those users to whom our model assigns the highest values of our latent experience score, while `novices' are assigned the lowest value; Figure \ref{fig:genreVexp} then shows the difference in product bias terms between the two groups.

This simple plot highlights some of our main findings with regard to user evolution: firstly, there is significant variation between the ratings of beginner and expert users, with the two groups differing by up to half a star in either direction. Secondly, there exist entire \emph{genres} of products that are preferred almost entirely by experts or by beginners: beginners give higher ratings to almost all lagers, while experts give higher ratings to almost all strong ales; thus we might conclude that strong ales are an `acquired taste'.

Finally, we find a strong correlation between the overall popularity of a product, and how much it is preferred by experienced users: experts tend to give the harshest ratings to the lowest-rated products, while they give the most generous ratings to the highest-rated products (continuing our analogy, they have learned to `fully appreciate' them). Thus while a lager such as \emph{Bud Light} is disliked by everybody, it is \emph{most disliked} by experts; one of the most popular beers in the entire corpus, \emph{Firestone XV}, is liked by everybody, but is \emph{most liked} by experts. We find such observations to be quite general across many datasets that we consider.

\subsection*{Contribution and Findings}
We propose a latent-factor model for the evolution of user experience in product recommendation systems, and compare it to other models of user and community evolution. While temporal dynamics and concept drift have been studied in online reviews~\cite{godes2012,koren10,moe12}, the critical difference between our model and existing work on temporal dynamics is how we treat experience as a function of time. Existing models for temporal dynamics work under the hypothesis that two users will respond most similarly to a product if they rate the product \emph{at the same time}. In contrast, we model the \emph{personal} evolution of users' tastes over time. We show that two users respond most similarly to a product if they review it \emph{at the same experience level}, even if there is significant separation in time between their reviews. Our model allows us to capture similarities between users, even when their reviews are temporally far apart.

In terms of experiments, we evaluate our model on five product rating websites, using novel corpora consisting of over 15 million reviews. Our ratings come from diverse sources including beers, wines, gourmet foods, and movies.

We find that our model of user \emph{experience} significantly outperforms traditional recommender systems, and similar alternatives that model user evolution at the level of products and communities.

Finally, we find that our latent experience parameters are themselves useful as \emph{features} to facilitate further study of how experts and novices behave. For example, we discover that experienced users rate top products more generously than beginners, and bottom products more harshly (as in Fig.~\ref{fig:genreVexp}); we find that users who fail to gain experience are likely to abandon the community; we find that experts' ratings are easier to predict than those of beginners; and we find that experienced users agree more closely when reviewing the same products. We can also use the notion of experience to discover which products, or categories of products, are preferred by beginners or by experts---that is, we can discover which products are \emph{acquired tastes}.

The rest of this paper is organized as follows: We describe our models for user evolution in Section \ref{sec:models}, and we describe how to train these models in Section \ref{sec:training}. In Section \ref{sec:experiments} we describe our novel rating datasets and evaluate our models. In Section \ref{sec:qualitative} we examine the role of our latent experience variables in detail, before reviewing related work in Section~\ref{sec:related} and concluding in Section \ref{sec:discussion}.

\section{Models of User Evolution}
\label{sec:models}

We design models to evaluate our hypothesis that `experience' is a critical underlying factor which causes users' ratings to evolve. We do so by considering alternate paradigms of user and community evolution, and determining which of them best fits our data.

\begin{figure}[t]
\vspace{1.4mm}
Community evolution at uniform intervals:
\begin{center}
\parbox[c]{0.08\columnwidth}{\textbf{(\modela)}}\parbox[c]{0.92\columnwidth}{\ \ \fbox{\includegraphics[width=0.85\columnwidth]{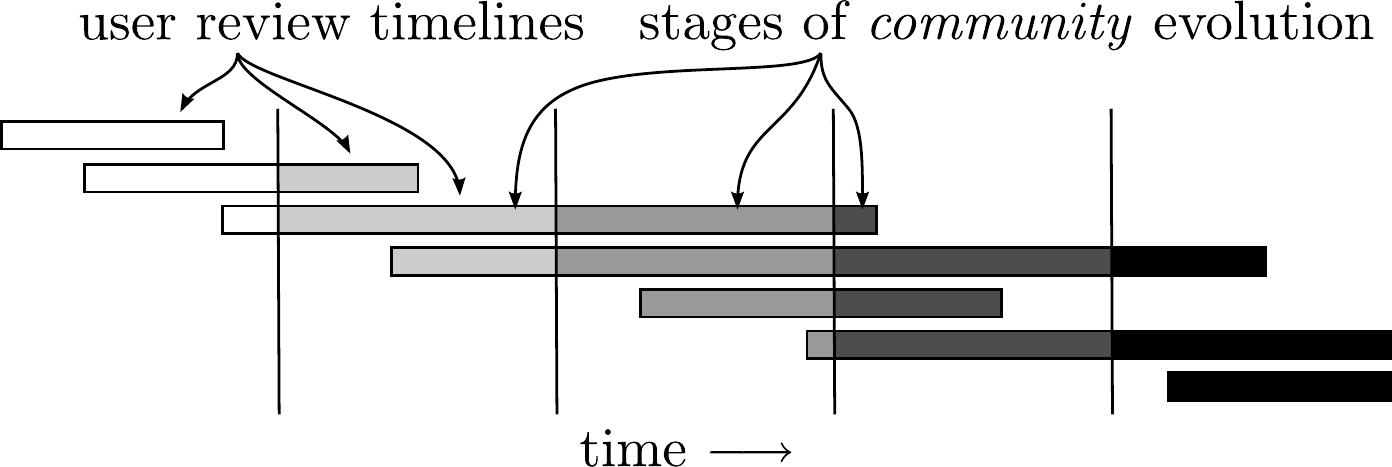}}}
\end{center}

Individual user evolution at uniform intervals:
\begin{center}
\parbox[c]{0.08\columnwidth}{\textbf{(\modelb)}}\parbox[c]{0.92\columnwidth}{\ \ \fbox{\includegraphics[width=0.85\columnwidth]{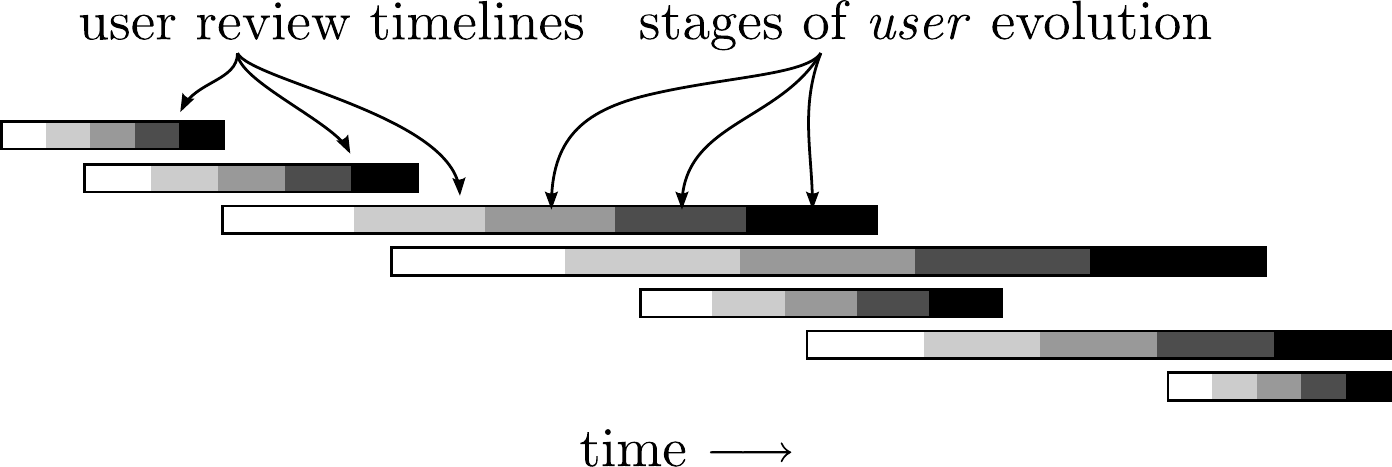}}}
\end{center}

Community evolution at learned intervals:
\begin{center}
\parbox[c]{0.08\columnwidth}{\textbf{(\modelc)}}\parbox[c]{0.92\columnwidth}{\ \ \fbox{\includegraphics[width=0.85\columnwidth]{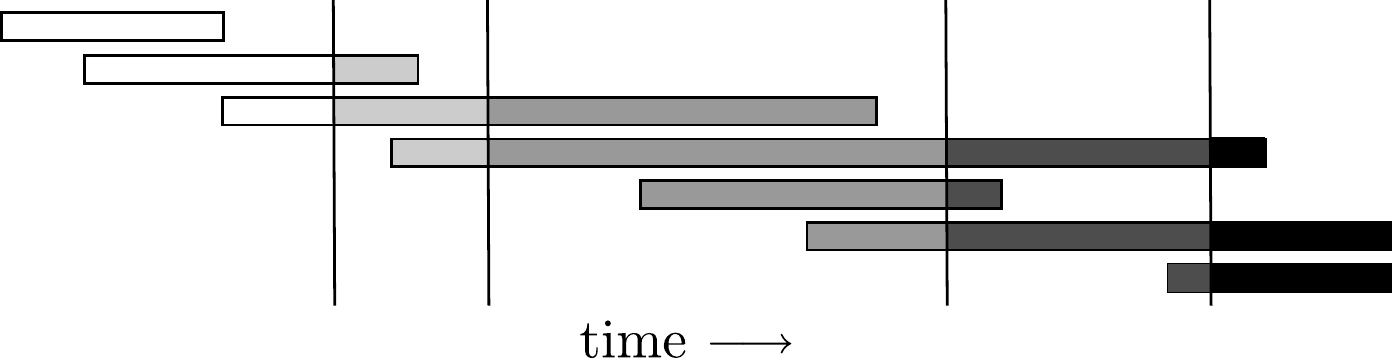}}}
\end{center}

Individual user evolution at learned intervals:
\begin{center}
\parbox[c]{0.08\columnwidth}{\textbf{(\modeld)}}\parbox[c]{0.92\columnwidth}{\ \ \fbox{\includegraphics[width=0.85\columnwidth]{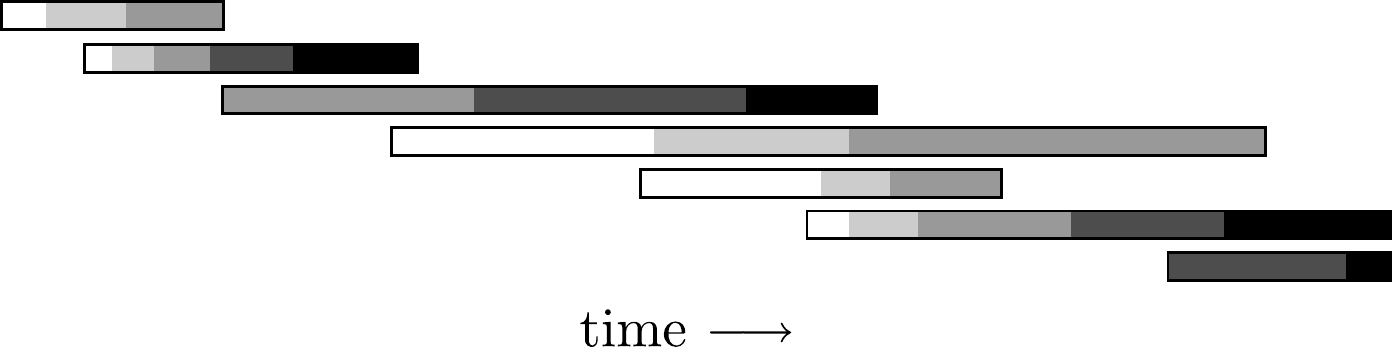}}}
\end{center}

\caption{Visualization of the models we consider. Horizontal bars represent user review timelines; colors within each bar represent evolution parameters for each user. \label{fig:models}}
\end{figure}

Figure \ref{fig:models} visualizes each of the four models we consider. Each horizontal bar represents a single user's review timeline, from their first to their last review; the color within each bar represents the evolution of that user or their community. At each of these stages of evolution, a different recommender system is used to estimate their ratings. Recommender systems for adjacent stages are regularized to be similar, so that transitions between successive stages are smooth.

\xxhdr{(a) Community evolution at uniform intervals:}
First we consider a model where `stages of evolution' appear at uniform time intervals throughout the history of the community. The model of Figure \ref{fig:models} (\modela) is in some sense the most similar to existing works \cite{koren10,xiang,xiong} that model evolution of users and products using a single global `clock'. The intuition behind this model is that communities evolve over time, and prefer different products at different time periods.

\xxhdr{(b) User evolution at uniform intervals:} 
We extend the idea of community evolution and apply it directly to individual users (Fig.~\ref{fig:models} (\modelb)). This model captures the intuition that users go through different life-stages or experience levels and their preferences then depend on their current life stage.
 
\xxhdr{(c) Community evolution at learned intervals:}
This model extends (\modela) by \emph{learning} the rates at which communities change over time (Fig.~\ref{fig:models} (\modelc)). The model is based on the intuition that a community may not evolve at a uniform rate over time and that it is worth modeling different stages of community evolution.

\xxhdr{(d) Individual user evolution at learned intervals:}
Last, we consider a model where each user individually progresses between experience levels at their own {\em personal rate} (Fig.~\ref{fig:models} (\modeld)). This model is the most expressive of all four and is able to capture interesting phenomena. For example, some users may become experts very quickly while others may never reach the highest level of experience; others may behave like experts from the time they join (e.g.~the bottom right user of Fig.~\ref{fig:models} (\modeld)). This model is able capture such types of behavior.

Models (\modela) and (\modelb) are designed to assess whether user evolution is guided by changes at the level of individual users, or by changes in the community at large. We find that, given enough data, both models lead to modest improvements over traditional latent-factor recommender systems, though there is no clear winner between the two. Once we learn the stages at which users and communities evolve, as in (\modelc) and (\modeld), we significantly outperform traditional recommender systems, though the benefit of learning is much higher when we model evolution at the level of each individual user, i.e., when we treat `evolution' as analogous to `becoming an expert'.

Put simply, we fit recommender systems for different stages of user evolution, and the models differ only in terms of how users \emph{progress} between stages. Thus the actual recommender systems used by each model have the same number of parameters, though the models of Figure \ref{fig:models} (\modelc) and (\modeld) have additional parameters that control \emph{when} users evolve. The model of Figure \ref{fig:models} (\modeld) is the most expressive, in the sense that it has enough flexibility to represent each of the other models. For example, if no evolution took place at the level of individual users, the model of Figure \ref{fig:models} (\modeld) could learn latent parameters similar to those of Figure \ref{fig:models} (\modelc). In practice, we find that this is not the case; rather we learn dynamics of individual users that are quite different to those at the level of communities.

\subsection*{Model Specification}
We shall first describe our most general model, namely that of Figure \ref{fig:models} (\modeld). The other models in Figure \ref{fig:models} can later be treated as special cases.

We start with the `standard' latent-factor recommender system \cite{handbook}, which predicts ratings for user/item pairs $(u,i)$ according to
\begin{equation*}
 \mathit{rec}(u,i) = \alpha + \beta_u + \beta_i + \langle \gamma_u, \gamma_i \rangle.
\label{eq:rec}
\end{equation*}
Here, $\alpha$ is a global offset, $\beta_u$ and $\beta_i$ are user and item biases (respectively), and $\gamma_u$ and $\gamma_i$ are latent user and item features.

Although this simple model can capture rich interactions between users and products, it ignores \emph{temporal} information completely, i.e., users' review histories are treated as unordered sets. Even so, such models yield excellent performance in practice \cite{Koren09}.

We wish to encode temporal information into such models via the proxy of \emph{user experience}. We do so by designing separate recommender systems for users who have different experience levels. Naturally, a user's experience is not fixed, but rather it evolves over time, as the user consumes (and rates) more and more products.

For each of a user's ratings $r_{ui}$, let $t_{ui}$ denote the time at which that review was entered (for simplicity we assume that each product is reviewed only once). For each of a user's ratings, we will fit a latent variable, $e_{ui}$, that represents the `experience level' of the user $u$, at the time $t_{ui}$.

Each user's experience level evolves over time. As a model assumption, we constrain each user's experience level to be a non-decreasing function of time. That is, a user never becomes \emph{less} experienced as they review additional products. We encode this using a simple monotonicity constraint on time and experience:
\begin{equation}
 \forall u,i,j\quad t_{ui} \geq t_{uj} \Rightarrow e_{ui} \geq e_{uj}.
\label{eq:constraint}
\end{equation}
What this constraint means from a modeling perspective is that different users evolve in similar ways to each other, regardless of the specific time they arrive in the community.

In practice, we model experience as a categorical variable that takes $E$ values, i.e., $e_{ui} \in \lbrace 1 \ldots E \rbrace$. Note that it is not required that a user achieves all experience levels: some users may already be experienced at the time of their first review, while others may fail to become experienced even after many reviews.

Assuming for the moment that each experience parameter $e_{ui}$ is observed, we proceed by fitting $E$ separate recommender systems to reviews written at different experience levels. That is, we train $\mathit{rec}_1(u,i) \ldots \mathit{rec}_E(u,i)$ so that each rating $r_{ui}$ is predicted using $\mathit{rec}_{e_{ui}}(u,i)$. As we show in Section \ref{sec:training}, we regularize the parameters of each of these recommender systems so that user and product parameters evolve `smoothly' between experience levels.

In short, each of the parameters of a standard recommender system is replaced by a parameter that is a function of experience:
\begin{multline}
 \mathit{rec}(u,i) = \mathit{rec}_{e_{ui}}(u,i)\\
                   = \alpha(e_{ui}) + \beta_u(e_{ui}) + \beta_i(e_{ui}) + \langle \gamma_u(e_{ui}), \gamma_i(e_{ui}) \rangle.
\label{eq:rec2}
\end{multline}
Such a model is quite general: rather than assuming that our model parameters evolve gradually over time, as in \cite{koren10}, we assume that they evolve gradually as a function of experience, which is \emph{itself} a function of time, but is learned \emph{per user}. Thus we are capable of learning whether users' `experience' parameters simply mimic the evolution of the entire community, as in Figure \ref{fig:models} (\modelc), or whether there are patterns of user evolution that occur independently of when they arrive in the community, as in Figure \ref{fig:models} (\modeld).

Because of this generality, all of the models from Figure \ref{fig:models} can be seen as special cases of \eq{eq:rec2}. Firstly, to model evolution of communities (rather than of individual users), we change the monotonicity constraint of \eq{eq:constraint} so that it constrains all reviews (rather than all reviews per user):
\begin{equation}
 \forall u,v,i,j\quad t_{ui} \geq t_{vj} \Rightarrow e_{ui} \geq e_{vj}.
 \label{eq:constraint2}
\end{equation}

Secondly, the `learned evolution' models (\modelc, \modeld) and the non-learned models (\modela, \modelb) differ in terms of how we fit the experience parameters $e_{ui}$. For the non-learned models, experience parameters are set using a fixed schedule: either they are placed at uniformly spaced time points throughout the entire corpus, as in Figure \ref{fig:models} (\modela), or they are placed at uniformly spaced time points for each individual user's history, as in Figure \ref{fig:models} (\modelb).

In the next section, we describe how to learn these parameters, i.e., to model the points at which a community or an individual user changes.

\section{Training the Models}
\label{sec:training}

We wish to optimize our model parameters and latent experience variables so as to minimize the mean-squared-error of predictions on some set of training ratings $r_{ui} \in \mathcal T$. Suppose each of our $E$ recommender systems has parameters $$\Theta_e = (\alpha(e); \beta_u(e); \beta_i(e); \gamma_u(e); \gamma_i(e)),$$ and that the set of all experience parameters $e_{ui}$ is denoted as $\mathcal E$. Then we wish to choose the optimal $(\hat{\Theta}, \hat{\mathcal E})$ according to the objective
\begin{multline}
 (\hat{\Theta}, \hat{\mathcal E}) = \argmin_{\Theta, \mathcal E} \sum_{r_{ui} \in \mathcal T} \frac{1}{|\mathcal T|} (\mathit{rec}_{e_{ui}}(u,i) - r_{ui})^2 + \lambda\Omega(\Theta)\\
\text{s.t.~} t_{ui} \geq t_{uj} \Rightarrow e_{ui} \geq e_{uj}.
\label{eq:train}
\end{multline}
This equation has three parts: the first is the mean-squared-error of predictions, which is the standard objective used to train recommender systems. The second part, $\Omega(\Theta)$, is a regularizer, which penalizes `complex' models $\Theta$. Assuming that there are $U$ users, $I$ items, $E$ experience levels, and $K$ latent factors, then our model has $(1 + U + I + U\cdot K + I\cdot K)\times E$ parameters, which will lead to overfitting if we are not careful. In practice, similar experience levels should have similar parameters, so we define $\Omega(\Theta)$ using the smoothness function
\begin{equation}
 \Omega(\Theta) = \sum_{e = 1}^{E - 1} \| \Theta_e - \Theta_{e+1} \|_2^2,
\end{equation}
where $||\cdot||_2^2$ is the squared $\ell_2$ norm. This penalizes abrupt changes between successive experience levels. $\lambda$ is a regularization hyperparameter, which `trades-off' the importance of regularization versus prediction accuracy at training time. We select $\lambda \in {10^0 \ldots 10^5}$ by withholding a fraction of our training data for validation, and choosing the value of $\lambda$ that minimizes the validation error. The third and final part of \eq{eq:train} is the constraint of \eq{eq:constraint}, which ensures that our latent experience parameters are monotonically non-decreasing for each user.

Simultaneously optimizing all of the parameters in \eq{eq:train} is a difficult problem, in particular it is certainly not convex \cite{Koren09}. We settle for a local optimum, and optimize the parameters $\Theta$ and $\mathcal E$ using coordinate ascent \cite{MacKay}.
That is, we alternately optimize \eq{eq:train} for $\Theta$ given $\mathcal E$, and for $\mathcal E$ given $\Theta$.

Optimizing \eq{eq:train} for $\Theta$ given $\mathcal E$, while itself still a non-convex problem, can be approached using standard techniques, since it essentially reduces to optimizing $E$ separate recommender systems. In practice we optimize model parameters during each iteration using L-BFGS \cite{lbfgs}, a quasi-Newton method for non-linear optimization of problems with many variables.

Alternately, optimizing \eq{eq:train} for $\mathcal E$ given $\Theta$ means assigning each of a user's reviews to a particular recommender system, corresponding to that review's experience level. The best assignment is the one that minimizes the mean-squared-error of the predictions, subject to the monotonicity constraint.

Optimizing a sequence of discrete variables subject to a monotonicity constraint can be solved efficiently using dynamic programming: it is related to the \emph{Longest Common Subsequence} problem \cite{lcs00}, which admits a solution whose running time (per user) is bilinear in $E$ and the number of ratings in their history.

\begin{figure}
\begin{center}
 \includegraphics[scale=1]{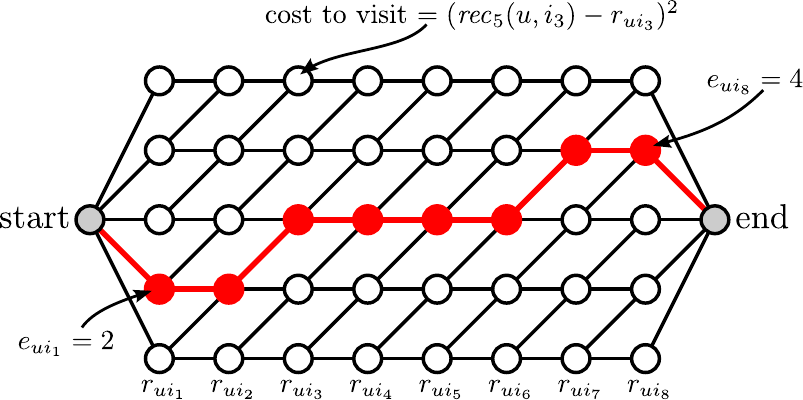}
\end{center}
\caption{Experience fitting as a dynamic programming problem. Rows represent experience levels, columns represent ratings, ordered by time. \label{fig:seq}}
\end{figure}

This procedure is visualized in Figure \ref{fig:seq}, for $E=5$ experience levels, and a user with 8 ratings. Rows represent each of the five experience levels, while columns represent each of a particular user's ratings, ordered by time. The optimal non-decreasing set of experience levels is the shortest path from the `start' to the `end' of this graph, where the cost of visiting a node with rating $r_{ui}$ at experience level $k$ is the prediction error $(\mathit{rec}_k(u,i) - r_{ui})^2$.

These two steps are repeated until convergence, that is, until $\mathcal E$ does not change between successive iterations. On our largest datasets, all parameters could be optimized in a few hours on a standard desktop machine.

Again, the above procedure refers to the most general version of our model, in which we learn monotonic evolution parameters \emph{per user}, as depicted in Figure \ref{fig:models} (\modeld). Training the community version of our model (Fig.~\ref{fig:models} (\modelc)) simply means replacing the monotonicity constraint of \eq{eq:constraint} with that of \eq{eq:constraint2}.

\section{Experiments}
\label{sec:experiments}

Our goal in this section is to evaluate the models described in Figure \ref{fig:models}. We compare the following models:
\begin{description}
 \item [\parbox{0.035\textwidth}{lf:}] A standard latent-factor recommender system \cite{korenSurvey}.
 \item [\parbox{0.035\textwidth}{\modela:}] A model whose parameters evolve for the entire community as a function of time (Fig.~\ref{fig:models} (\modela)).
 \item [\parbox{0.035\textwidth}{\modelb:}] A model whose parameters evolve independently for each user (Fig.~\ref{fig:models} (\modelb)).
 \item [\parbox{0.035\textwidth}{\modelc:}] A model whose parameters evolve for the entire community as a function of time, where the `stages' of evolution are learned (Fig.~\ref{fig:models} (\modelc)).
 \item [\parbox{0.035\textwidth}{\modeld:}] A model whose parameters evolve independently for each user, where the stages of evolution are learned (Fig.~\ref{fig:models} (\modeld)).
\end{description}

The models of Figure \ref{fig:models} (\modela) and (\modelc) are most similar to existing models for temporal evolution, e.g.~\cite{Koren09}: item parameters are shared by ratings made at the same time. We aim to compare this to models where parameters are shared by users at the same experience level, regardless of the specific time they arrive in the community (as in Fig.~\ref{fig:models} (\modeld)).

\subsection*{Experimental Setup}
To evaluate each method, we report the Mean Squared Error (MSE) on a fraction of our data withheld for testing, that is, for our test set $\mathcal{U}$ we report
\begin{equation}
 \text{MSE}(\mathcal U) = \frac{1}{|\mathcal U|} \sum_{r_{ui} \in \mathcal U} (\mathit{rec}_{e_{ui}}(u,i) - r_{ui})^2.
\end{equation}

We also use a validation set of the same size to choose the hyperparameter $\lambda$. Throughout our experiments we set the number of experience levels, and the number of latent product and item dimensions to $E = 5$ and $K = 5$; larger values did not significantly improve performance in our experience.

Since it is unlikely to be fruitful to model the evolution of users who have rated only a few products, we compare our models on users with at least 50 ratings. Users with fewer than 50 ratings are not discarded, but rather their ratings are combined so that they are treated using a single `background' model; we then model the evolution of this entire group as though it were a single user.

We use two schemes to build our test sets:
our first scheme consists of selecting a \emph{random} sample of reviews from each user. This is the standard way of selecting test data for `flat' models that do not model temporal dynamics.
The second scheme we use to build our test set is to consider the \emph{final} reviews for each user.

The latter setting represents how such a system would be used in practice, in the sense that our goal is to predict how users would respond to a product \emph{now}, rather than to make \emph{post hoc} predictions about how they \emph{would have} responded in the past. However, sampling reviews in this way biases our test set towards reviews written by more experienced users; this is no longer the case when we sample reviews randomly.

Of course, we do not fit latent experience parameters for the ratings in our test set. Thus for each rating used for testing, we assign it the experience level of its chronologically nearest training rating.

\begin{table}
\begin{center}
\begin{tabular}{lrrr}
\multicolumn{1}{c}{\bf dataset} & \multicolumn{1}{c}{\hspace{-2mm}\bf \#users} & \multicolumn{1}{c}{\bf \#items} & \multicolumn{1}{c}{\bf \#ratings}\\
\hline\\[-2.7mm]
Beer (beeradvocate) & 33,387 & 66,051 & 1,586,259\\
Beer (ratebeer) & 40,213 & 110,419 & 2,924,127\\
Fine Foods (amazon) & 218,418 & 74,442 & 568,454\\
Movies (amazon) & 759,899 & 267,320 & 7,911,684\\
Wine (cellartracker) & 44,268 & 485,179 & 2,025,995\\
\hline\\[-2.7mm]
TOTAL & 1,096,185 & 1,003,411 & 15,016,519
\end{tabular}
\normalsize
\end{center}
\caption{Dataset statistics. \label{tab:datasets}}
\end{table}

\subsection*{Datasets}
Our choice of rating data reflects a variety of settings where users are likely to have `acquired tastes'. The datasets we consider are summarized in Table \ref{tab:datasets}. Each of our datasets were obtained from public sources on the web using a crawler, and are made available for others to use.\footnote{\texttt{http://snap.stanford.edu/data/}} We consider the beer review websites \emph{BeerAdvocate} and \emph{RateBeer}, the wine review website \emph{CellarTracker}, as well as reviews from the \emph{Fine Foods} and \emph{Movies} categories from \emph{Amazon}. In total we obtain over 15 million ratings from these sources. In principle we obtain the \emph{complete} set of reviews from each of these sources; data in each of our corpora spans at least 10 years.

We previously considered \emph{BeerAdvocate} and \emph{RateBeer} data in \cite{palelager}, though not in the context of recommendation. Recommendation on (different) \emph{Amazon} data has been discussed in \cite{Jindal08opinionspam} and \cite{Linden03}.

Since each of these rating datasets has a different scale (e.g.~beers on \emph{RateBeer} are rated out of 20, wines on \emph{CellarTracker} are rated out of 100, etc.), before computing the MSE we first normalize all ratings to be on the scale $(0,5]$.

\subsection*{Evaluation}

\begin{table*}
\begin{center}
\begin{tabular}{l|@{\hspace{1mm}} >{\centering\arraybackslash}m{0.098\textwidth}@{\hspace{1mm}} >{\centering\arraybackslash}m{0.098\textwidth}@{\hspace{1mm}} >{\centering\arraybackslash}m{0.098\textwidth}@{\hspace{1mm}} >{\centering\arraybackslash}m{0.098\textwidth}@{\hspace{1mm}} >{\centering\arraybackslash}m{0.098\textwidth}@{\hspace{1mm}} >{\centering\arraybackslash}m{0.098\textwidth}@{\hspace{1mm}} >{\centering\arraybackslash}m{0.098\textwidth}}
& BeerAdv. (overall) & BeerAdv. (taste) & BeerAdv. (look) & RateBeer (overall) & Amazon Fine Foods & Amazon Movies & CellarTracker \\
\hline\\[-2mm]
\textbf{(lf)} latent-factor model & 0.452 (.01) & 0.442 (.01) & 0.313 (.01) & 0.496 (.01) & 1.582 (.02) & 1.379 (.00) & 0.055 (.00) \\
\textbf{(\modela)} community at uniform rate & 0.427 (.01) & 0.417 (.01) & 0.293 (.01) & 0.458 (.01) & 1.527 (.02) & 1.371 (.01) & 0.051 (.00) \\
\textbf{(\modelb)} user at uniform rate & 0.437 (.01) & 0.423 (.01) & 0.300 (.01) & 0.477 (.01) & 1.548 (.02) & 1.376 (.01) & 0.053 (.00) \\
\textbf{(\modelc)} community at learned rate & 0.427 (.01) & 0.417 (.01) & 0.293 (.01) & 0.458 (.01) & 1.529 (.02) & 1.371 (.01) & 0.051 (.00) \\
\textbf{(\modeld)} user at learned rate & \textbf{ 0.400 (.01) } & \textbf{ 0.399 (.01) } & \textbf{ 0.275 (.01) } & \textbf{ 0.406 (.01) } & \textbf{ 1.475 (.03) } & \textbf{ 1.051 (.01) } & \textbf{ 0.045 (.00) } \\
benefit of \textbf{(\modeld)} over \textbf{(lf)}
& 11.62\% & 9.73\% & 12.19\% & 18.26\% & 6.79\% & 23.80\% & 18.50\% \\
benefit of \textbf{(\modeld)} over \textbf{(\modelc)}
& 6.48\% & 4.12\% & 6.13\% & 11.42\% & 3.53\% & 23.34\% & 13.20\% \\
\end{tabular}
\normalsize
\vspace{-1.38mm}
\end{center}
\caption{Results on users' most recent reviews. MSE and standard error.\label{tab:resultsFinal}}
\end{table*}

\begin{table*}
\begin{center}
\begin{tabular}{l|@{\hspace{1mm}} >{\centering\arraybackslash}m{0.098\textwidth}@{\hspace{1mm}} >{\centering\arraybackslash}m{0.098\textwidth}@{\hspace{1mm}} >{\centering\arraybackslash}m{0.098\textwidth}@{\hspace{1mm}} >{\centering\arraybackslash}m{0.098\textwidth}@{\hspace{1mm}} >{\centering\arraybackslash}m{0.098\textwidth}@{\hspace{1mm}} >{\centering\arraybackslash}m{0.098\textwidth}@{\hspace{1mm}} >{\centering\arraybackslash}m{0.098\textwidth}}
& BeerAdv. (overall) & BeerAdv. (taste) & BeerAdv. (look) & RateBeer (overall) & Amazon Fine Foods & Amazon Movies & CellarTracker \\
\hline\\[-2mm]
\textbf{(lf)} latent-factor model & 0.430 (.01) & 0.408 (.01) & 0.319 (.01) & 0.492 (.01) & 1.425 (.02) & 1.099 (.01) & 0.049 (.00) \\
\textbf{(\modela)} community at uniform rate & 0.415 (.01) & 0.387 (.01) & 0.298 (.01) & 0.463 (.01) & 1.382 (.02) & 1.082 (.01) & 0.048 (.00) \\
\textbf{(\modelb)} user at uniform rate & 0.419 (.01) & 0.395 (.01) & 0.305 (.01) & 0.461 (.01) & 1.383 (.02) & 1.088 (.01) & 0.048 (.00) \\
\textbf{(\modelc)} community at learned rate & 0.415 (.01) & 0.386 (.01) & 0.298 (.01) & 0.461 (.01) & 1.374 (.02) & 1.082 (.01) & 0.048 (.00) \\
\textbf{(\modeld)} user at learned rate & \textbf{ 0.409 (.01) } & \textbf{ 0.373 (.01) } & \textbf{ 0.276 (.01) } & \textbf{ 0.394 (.01) } & \textbf{ 1.189 (.03) } & \textbf{ 0.711 (.01) } & \textbf{ 0.039 (.00) } \\
benefit of \textbf{(\modeld)} over \textbf{(lf)}
& 5.05\% & 8.61\% & 13.45\% & 19.94\% & 16.61\% & 35.31\% & 20.49\% \\
benefit of \textbf{(\modeld)} over \textbf{(\modelc)}
& 1.55\% & 3.33\% & 7.20\% & 14.50\% & 13.47\% & 34.32\% & 18.23\% \\
\end{tabular}
\normalsize
\vspace{-1.38mm}
\end{center}
\caption{Results on randomly sampled reviews. MSE and standard error.\label{tab:resultsRandom}}
\end{table*}

Results in terms of the Mean Squared Error (MSE) are shown in Tables \ref{tab:resultsFinal} and \ref{tab:resultsRandom}. Table \ref{tab:resultsFinal} shows the MSE on a test set consisting of the \emph{most recent} reviews for each user, while Table \ref{tab:resultsRandom} shows the MSE on a \emph{random} subset of users' reviews.

Table \ref{tab:resultsFinal} shows that our model significantly outperforms alternatives on all datasets. On average, it achieves a 14\% reduction in MSE compared to a standard latent factor recommender system, and a 10\% reduction compared to its nearest competitor, which models user evolution as a process that takes place at the level of entire communities. Note that due to the large size of our datasets, all reported improvements are significant at the 1\% level or better.

By considering only users' most recent reviews, our evaluation may be biased towards reviews written at a high level of experience. To address this possibility, in Table \ref{tab:resultsRandom} we perform the same evaluation on a \emph{random} subset of reviews for each user. Again, our model significantly outperforms all baselines. Here we reduce the MSE of a standard recommender system by 17\%, and the nearest competitor by 13\% on average.

Reviews from \emph{BeerAdvocate} and \emph{RateBeer} have multiple dimensions, or `aspects' to users' evaluations. Specifically, users evaluate beers in terms of their `taste', `smell', `look', and `feel' in addition to their overall rating \cite{palelager}. Tables \ref{tab:resultsFinal} and \ref{tab:resultsRandom} show results for two such aspects from \emph{BeerAdvocate}, showing that we obtain similar benefits by modeling users' evolution with respect to such aspects. Similar results from \emph{RateBeer} are omitted.

It is perhaps surprising that we gain the most significant benefits on movie data, and the least significant benefits on beer data. However, we should not conclude that movies require more expertise than beer: a more likely explanation is that our movie data has a larger \emph{spectrum} of expertise levels, whereas users who decide to participate on a beer-rating website are likely to already be somewhat `expert'.

We gain the most significant benefits when considering reviews written at all experience levels (as in Table \ref{tab:resultsRandom}) rather than considering users' most recent reviews (as in Table \ref{tab:resultsFinal}). However we should not conclude from this that experts are unpredictable (indeed in Section \ref{sec:qualitative} we confirm that experts are the \emph{most} predictable). Rather, since inexperienced users are \emph{less} predictable, we gain the most benefit by explicitly modeling them.

We also note that while we obtain significant benefits on \emph{Amazon} data, the mean-squared-errors for this dataset are by far the highest. One reason is that \emph{Amazon} users use a full spectrum of ratings from 1 to 5 stars, whereas \emph{CellarTracker} users (for example) rate wines on a smaller spectrum (after normalizing the ratings, most wines have scores above 4.25); this naturally leads to higher MSEs. Another reason is that our \emph{Amazon} data has many products and users with only a few reviews, so that we cannot do much better than simply modeling their bias terms. As we see in Section \ref{sec:qualitative}, bias terms differ significantly between beginners and experts, so that modeling expertise proves extremely beneficial on such data.

\section{Qualitative Analysis}
\label{sec:qualitative}

So far, we have used our models of user expertise to predict users' ratings, by fitting latent `experience' parameters to all ratings in our corpora. Now we move on to examine the role of these latent variables in more detail.

Throughout this section we use the term `expert' to refer to those reviewers (and ratings) that are assigned the highest experience level by our model (i.e., $e_{ui} = E$). We use the term `beginner' to refer to reviewers and ratings assigned the lowest level (i.e., $e_{ui} = 1$). Again, we acknowledge that `expertise' is an \emph{interpretation} of our model's latent parameters, and other interpretations may also be valid. However, in this section, we demonstrate that our latent parameters do indeed behave in a way that is consistent with our intuitive notion of expertise.

We begin by examining how a user's experience level impacts our ability to predict their rating behavior. Table \ref{tab:eVsMse} compares the prediction accuracy of our model on reviews written at different experience levels. We find in all but one case that users at the highest experience level have the lowest MSE (for \emph{Amazon Movies} they have the second lowest by a small margin). In other words their rating behavior can be most accurately predicted by our model. This is not to say that experts agree with \emph{each other} (which we discuss later); rather, it says that \emph{individual} experts are easier to model than other categories of user. Indeed, one can argue that some notion of `predictability' is a necessary condition for such users to be considered `experts' \cite{einhorn1974expert}.

While we find that beginners and intermediate users have lower prediction accuracy, it is surprisingly the `almost experts' ($e_{ui} = E - 1$) who are the \emph{least} predictable; from Table \ref{tab:eVsMse} we see that such users have the \emph{highest} MSE in three out of five cases. From this we might argue that users do not become experts via a smooth progression, but rather their evolution consists of several distinct stages.

\begin{table}
\begin{center}
\begin{tabular}{l|@{\hspace{2mm}} >{\centering\arraybackslash}c@{\hspace{2mm}} >{\centering\arraybackslash}c@{\hspace{2mm}} >{\centering\arraybackslash}c@{\hspace{2mm}} >{\centering\arraybackslash}c@{\hspace{2mm}} >{\centering\arraybackslash}c}
 & $e = 1$ & $e = 2$ & $e = 3$ & $e = 4$ & $e = 5$\\
\hline\\[-2mm]
BeerAdvocate & 0.423 & 0.396 & 0.471 & 0.449 & \textbf{0.358} \\
RateBeer & 0.494 & 0.469 & 0.408 & 0.533 & \textbf{0.300} \\
Amazon Fine Foods & 1.016 & 1.914 & 1.094 & 2.251 & \textbf{0.960} \\
Amazon Movies & 0.688 & \textbf{0.620} & 0.685 & 1.062 & 0.675 \\
CellarTracker & 0.061 & 0.039 & 0.041 & 0.037 & \textbf{0.028} \\
\end{tabular}
\normalsize
\end{center}
\caption{MSE per experience level $e$ \label{tab:eVsMse}}
\end{table}

\subsection*{Experience Progression}
\begin{figure*}[t]
\begin{center}
\hspace{5mm}\includegraphics[scale=0.555]{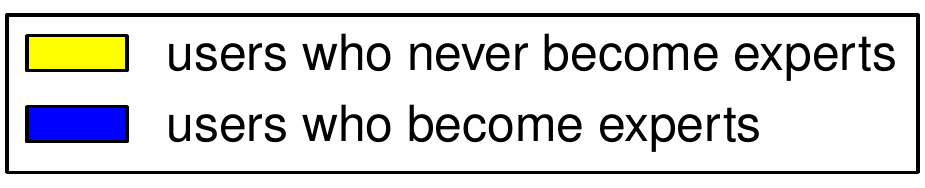}

\includegraphics[scale=0.555]{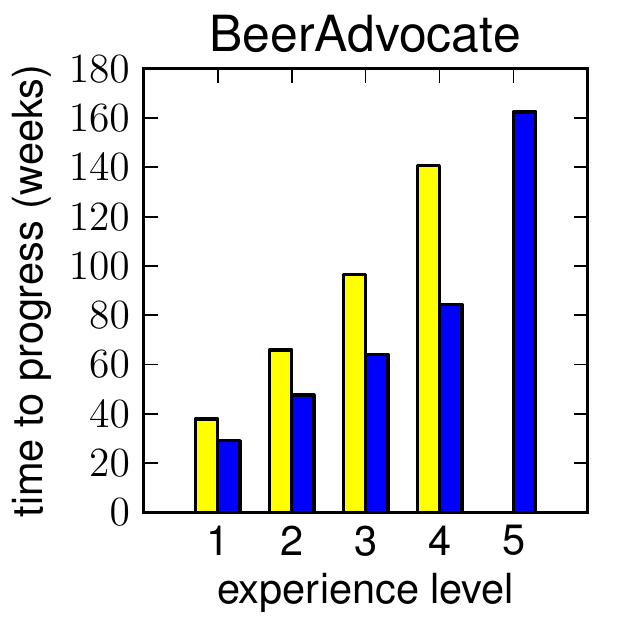}%
\includegraphics[scale=0.555]{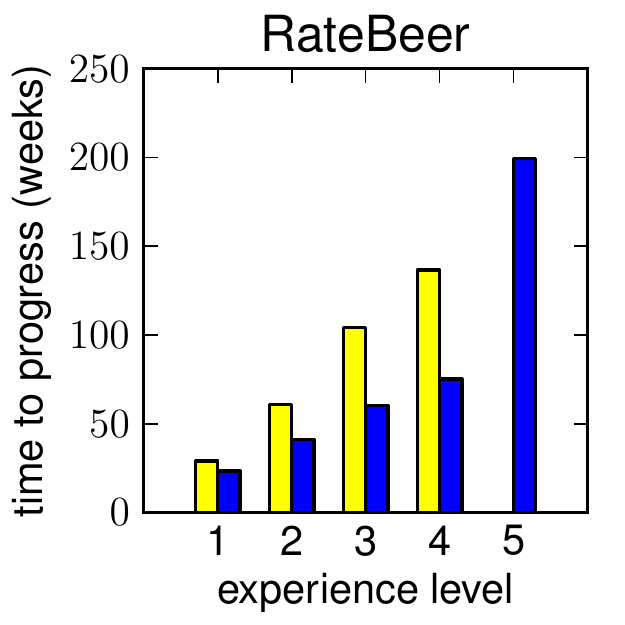}%
\includegraphics[scale=0.555]{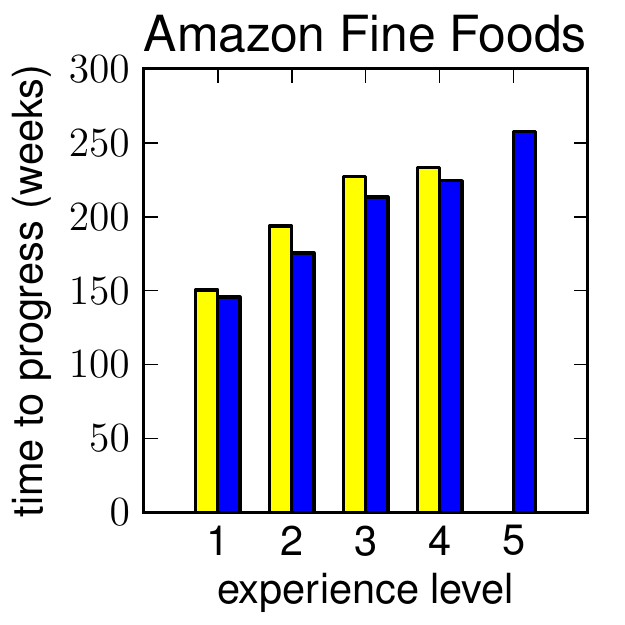}%
\includegraphics[scale=0.555]{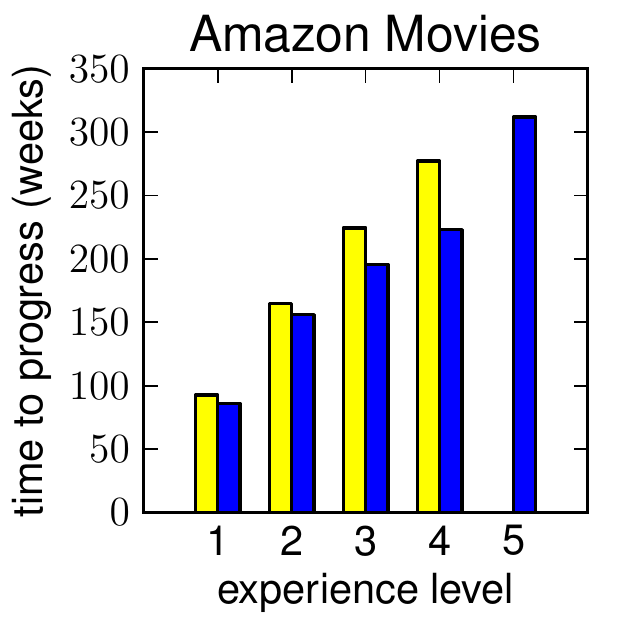}%
\includegraphics[scale=0.555]{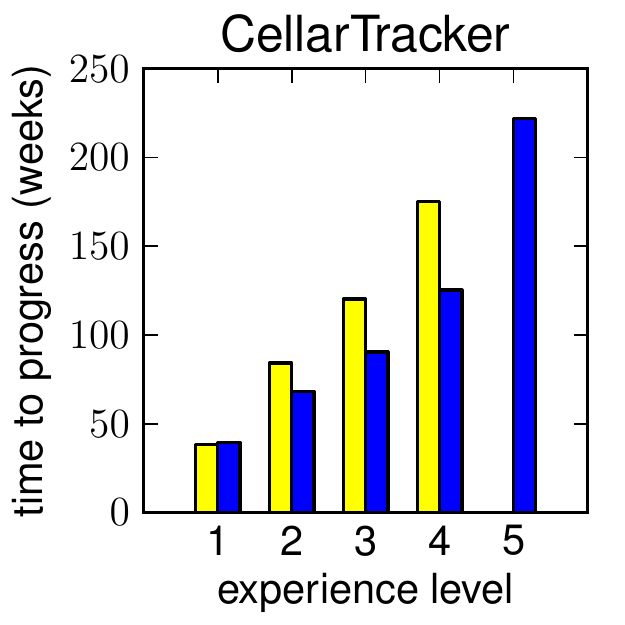}%

\includegraphics[scale=0.555]{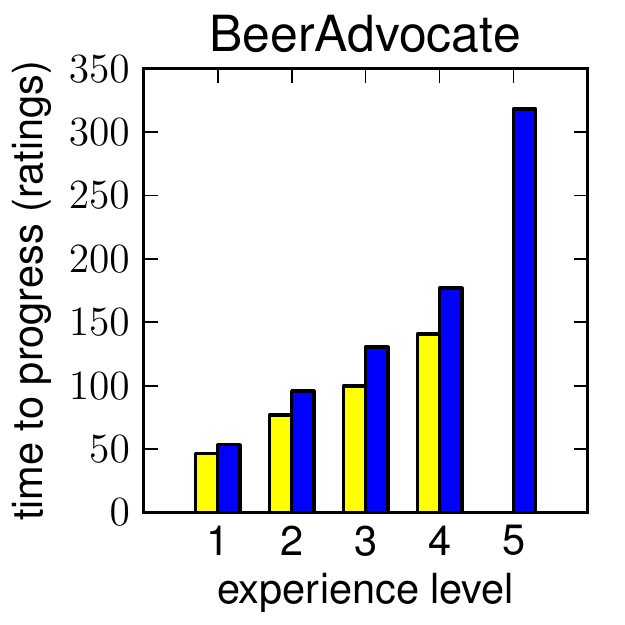}%
\includegraphics[scale=0.555]{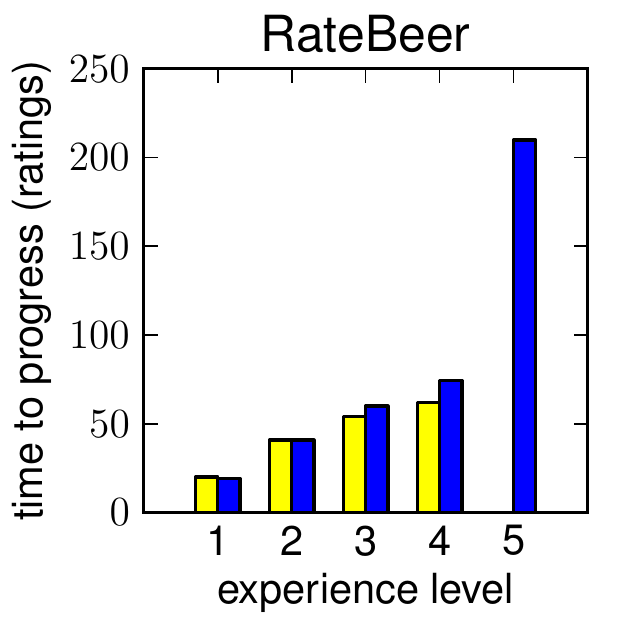}%
\includegraphics[scale=0.555]{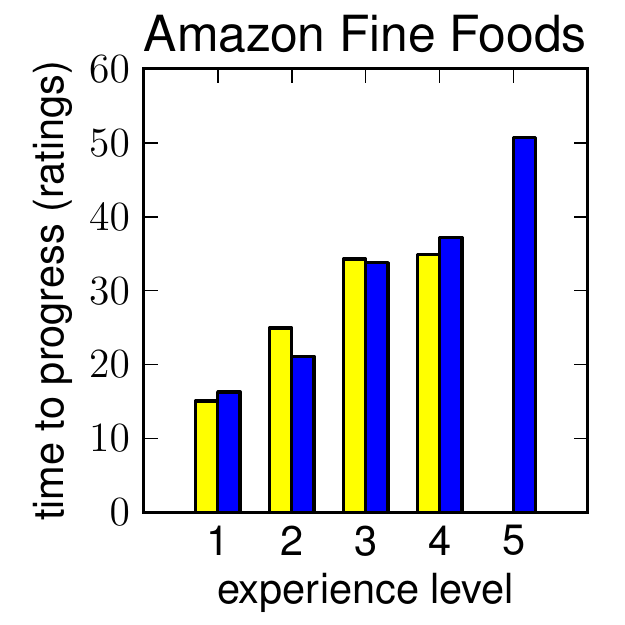}%
\includegraphics[scale=0.555]{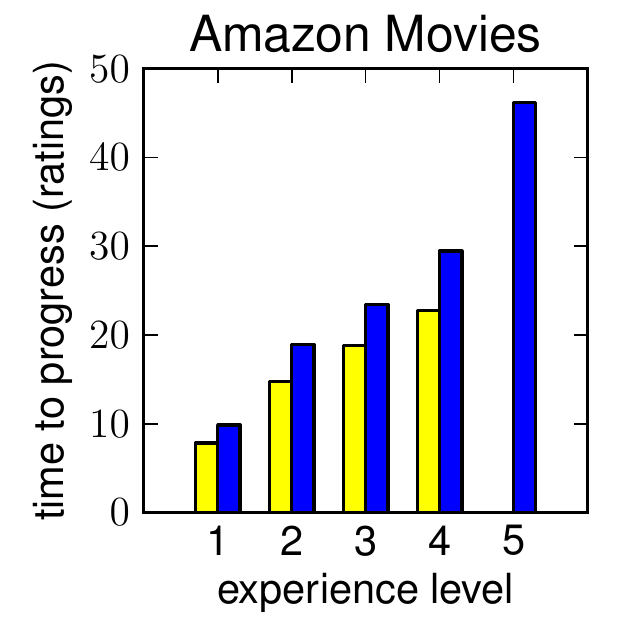}%
\includegraphics[scale=0.555]{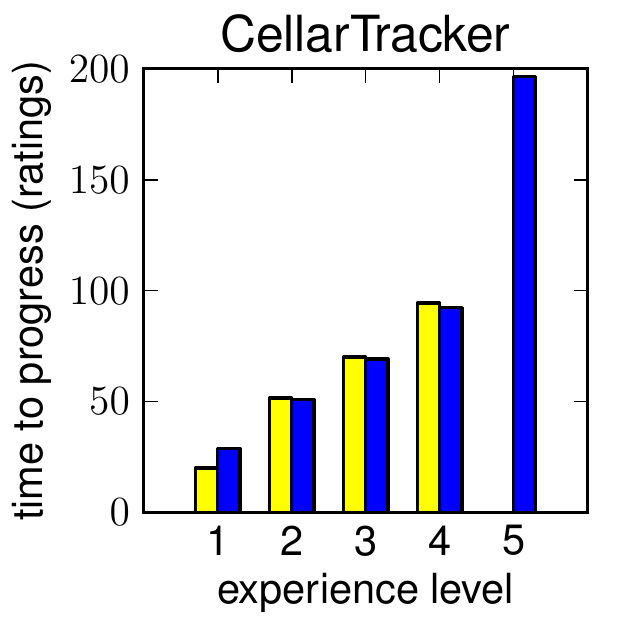}%

\vspace{-2.55mm}
\end{center}
\caption{Users who never become `experts' tend to progress slower than users who do. Cumulative time (top), and number of ratings (bottom), taken to progress between experience levels. \label{fig:times}}
\end{figure*}

Next, we study how users progress through experience levels as a function of time. Figure \ref{fig:times} shows the (cumulative) time taken to progress between experience levels (the final bar represents the entire lifetime of the user, since there is no further level to progress to). The dark blue bars show the progression for those users who progress through all levels of experience, i.e., it ignores those users who arrive to the site already experienced as well as those who never obtain the highest experience levels. The yellow bars show users who reach all but the highest experience level.

\xhdr{How much time is spent at each experience level?} First, we observe that on most datasets, the final experience level is the `longest', i.e., it covers the longest time period, and includes the largest number of reviews. This makes sense from the modeling perspective, when taken together with our previous finding that experts' ratings are easier to predict: the model is `finer-grained' during the stages of user evolution that are most difficult to fit accurately. Fewer distinct experience levels are required later on, once users' rating behavior has `converged'.

\xhdr{Do users who become experts differ from those who don't?} Secondly, Figure \ref{fig:times} compares users who progress through all levels of experience to users who do not. Yellow bars show the progression of users who reach all but the final experience level. Surprisingly, while such users enter roughly the same number of ratings per level (Fig.~\ref{fig:times}, bottom) as those users who eventually become experts, they do so much slower (Fig.~\ref{fig:times}, top). Thus it appears as though the \emph{rate} at which users write reviews, and not just the number of reviews they write, is tied to their progression.

\begin{figure*}[t]
\begin{center}
\includegraphics[scale=0.555]{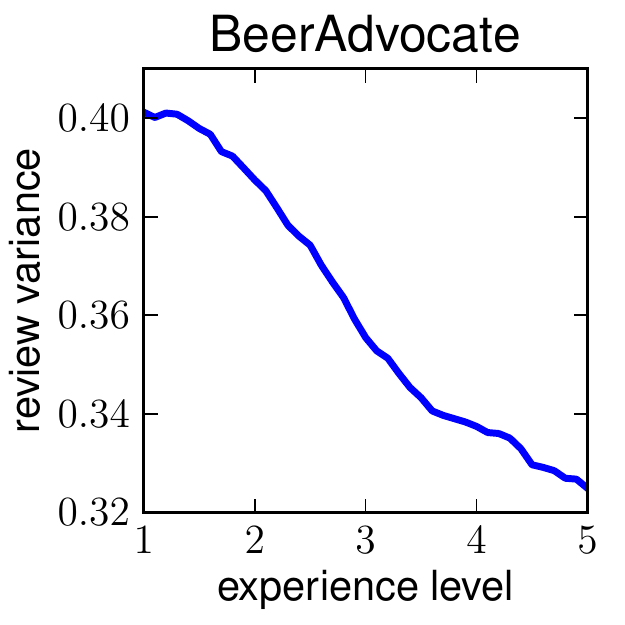}%
\includegraphics[scale=0.555]{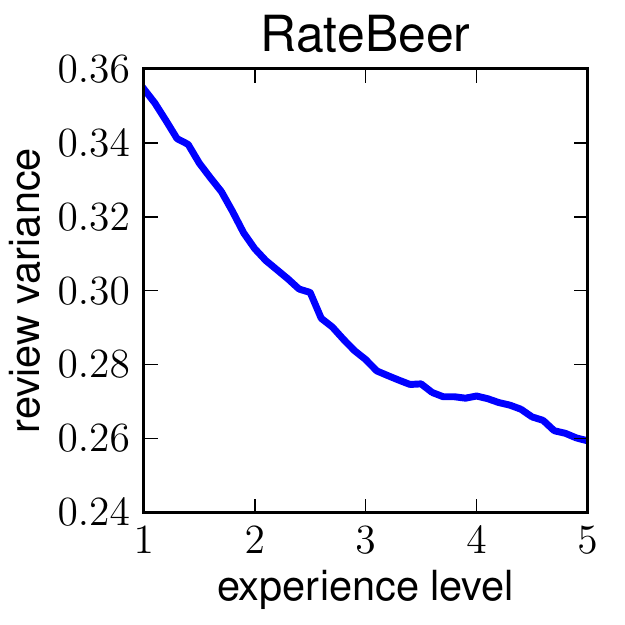}%
\includegraphics[scale=0.555]{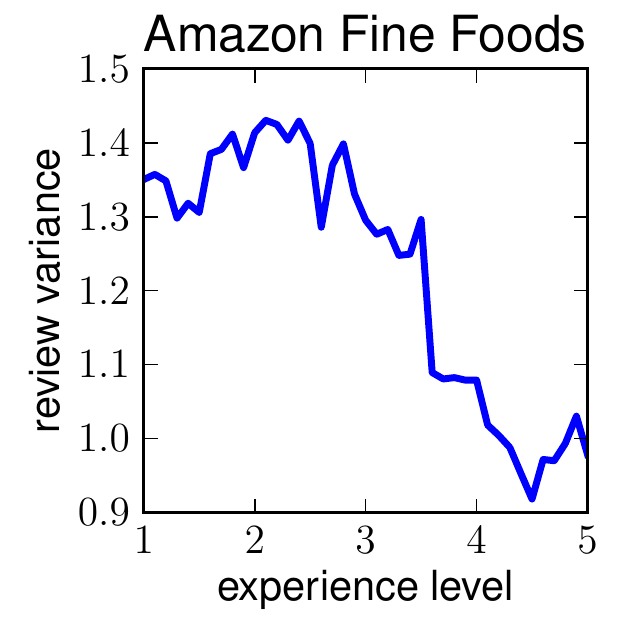}%
\includegraphics[scale=0.555]{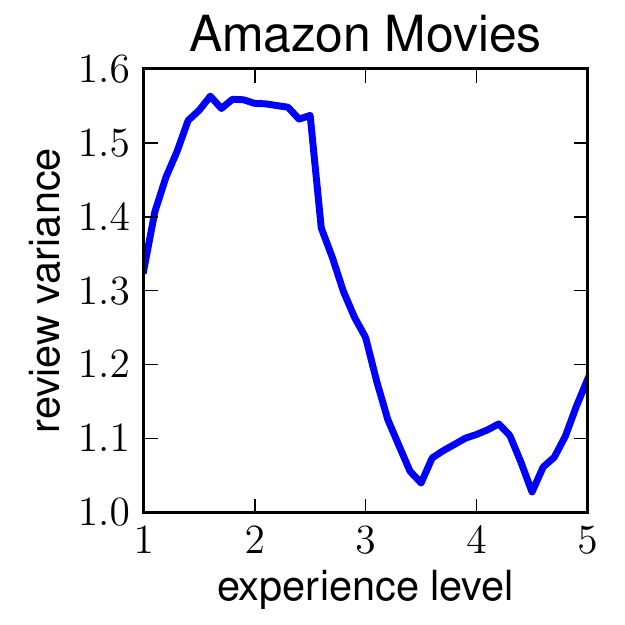}%
\includegraphics[scale=0.555]{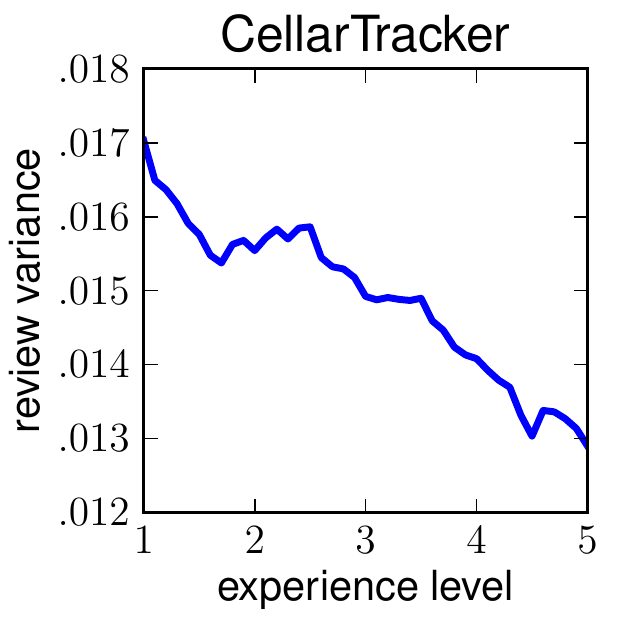}%

\vspace{-2.55mm}
\end{center}
\caption{Experienced users agree more about their ratings than beginners. Experience versus rating variance (when rating the same product). \label{fig:agreement}}
\end{figure*}

\xhdr{Do experts agree with each other?} Thirdly, Figure \ref{fig:agreement} shows the extent to which users \emph{agree} with each other as they become more experienced. `Agreement' has been argued to be another necessary condition to define users as experts \cite{einhorn1974expert}. To study this, we consider ratings \emph{of the same product, written at the same experience level}. Specifically, for each item $i$ and experience level $k$, we find the set of users who rated that item at that experience level, i.e., we find all $u$ such that $e_{ui} = k$. We then compute the variance of such ratings for every item and experience level. Our goal is to assess how this quantity changes as a function of users' experience. We do so for all products that were reviewed at least 5 times at the same experience level. Since this limits the amount of data we have to work with, we first linearly interpolate each user's experience function over time (so that their experience function is a piecewise linear function, rather than a step function), and compute this quantity across a sliding window.

Indeed, in Figure \ref{fig:agreement} we find that users do tend to agree with each other more as they become more experienced, i.e., their ratings have lower variance when they review the same products. This is consistent with our finding that experts' ratings are easier to predict than those of beginners.

\subsection*{User Retention}
\begin{figure*}[t]
\begin{center}
\hspace{5mm}\includegraphics[scale=0.555]{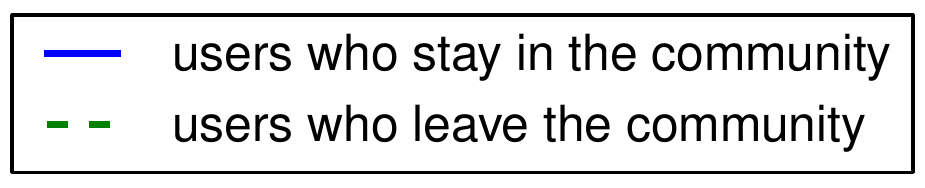}

\includegraphics[scale=0.555]{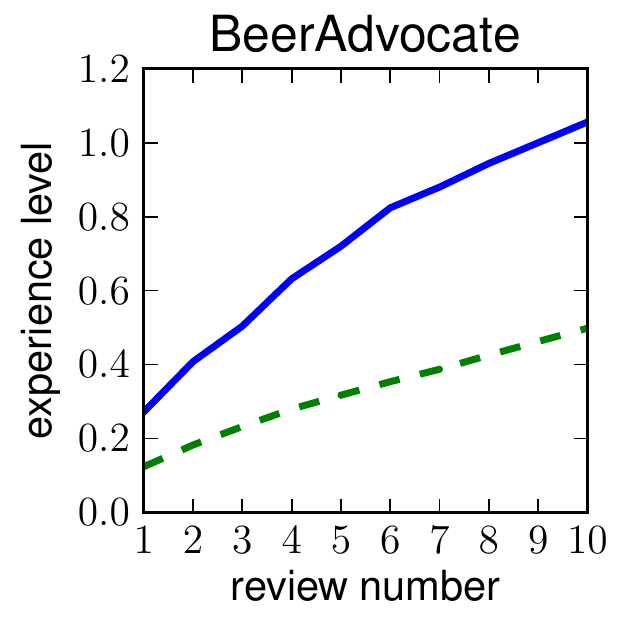}%
\includegraphics[scale=0.555]{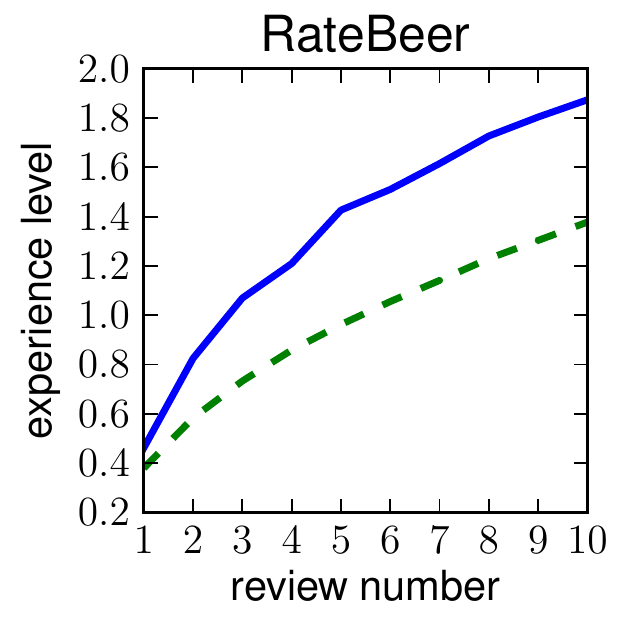}%
\includegraphics[scale=0.555]{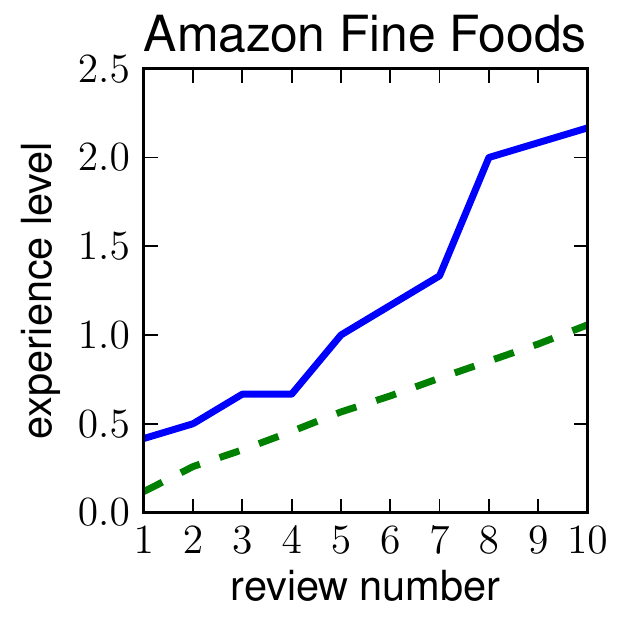}%
\includegraphics[scale=0.555]{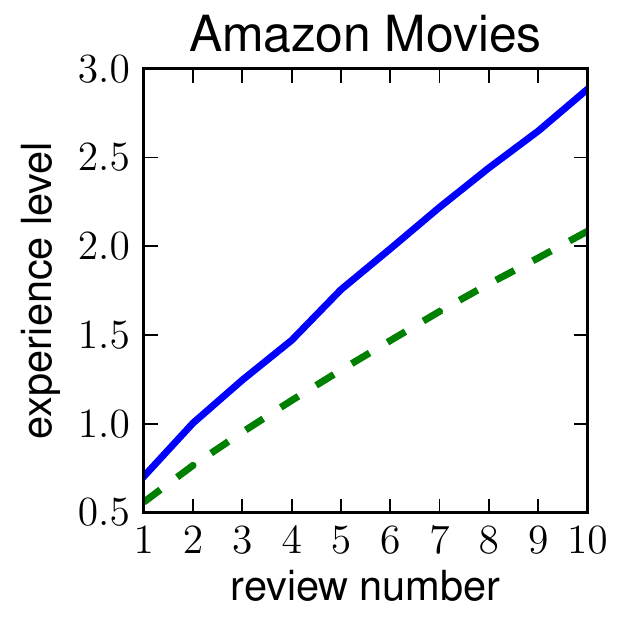}%
\includegraphics[scale=0.555]{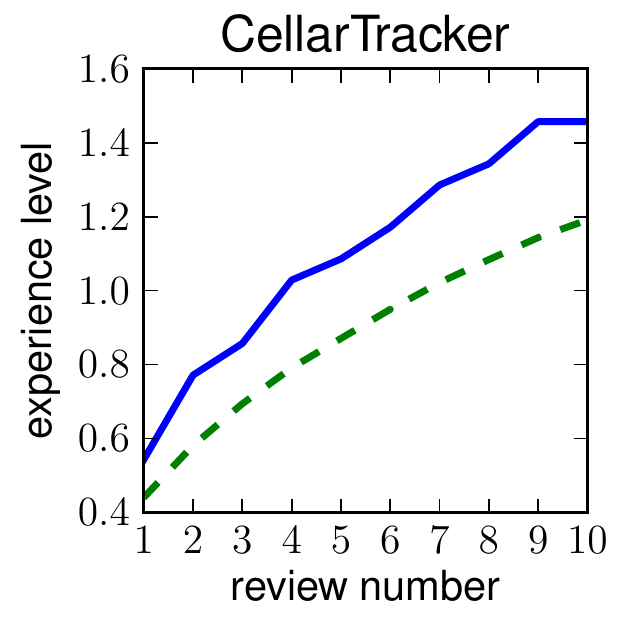}%

\includegraphics[scale=0.555]{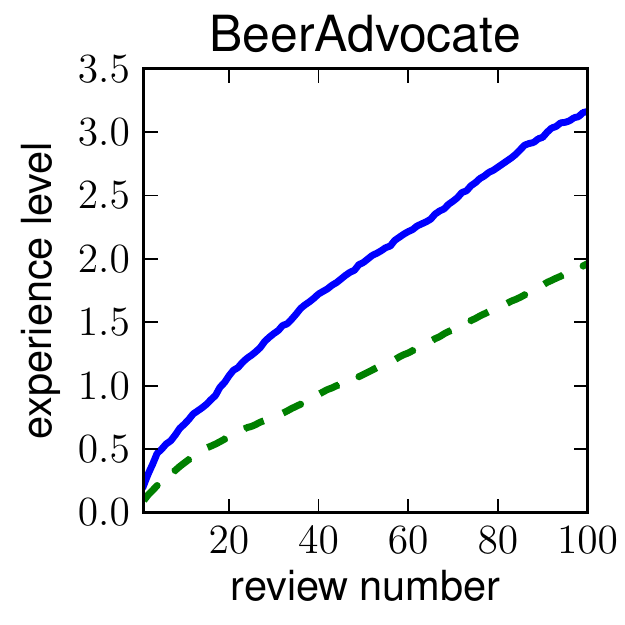}%
\includegraphics[scale=0.555]{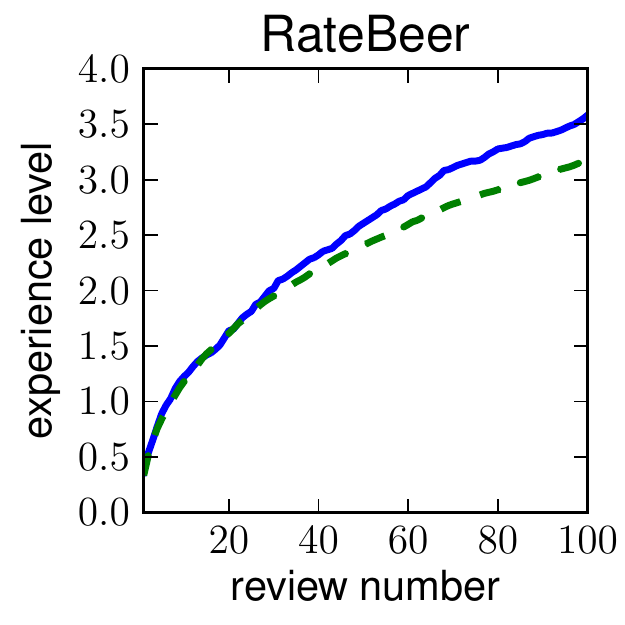}%
\includegraphics[scale=0.555]{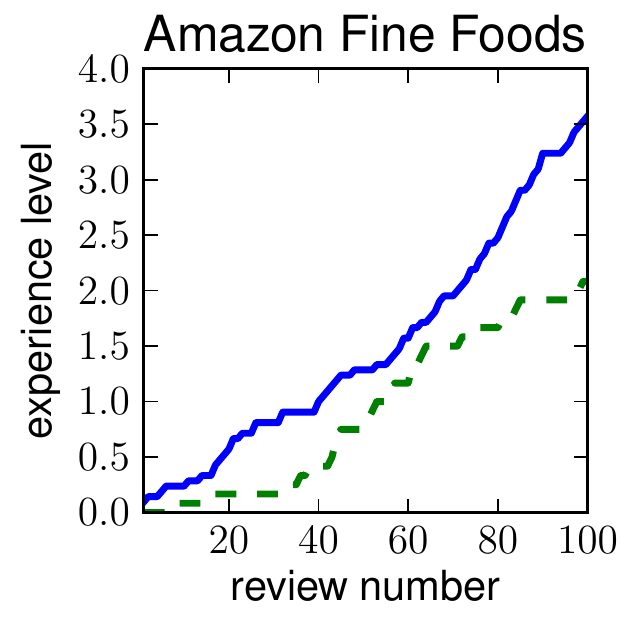}%
\includegraphics[scale=0.555]{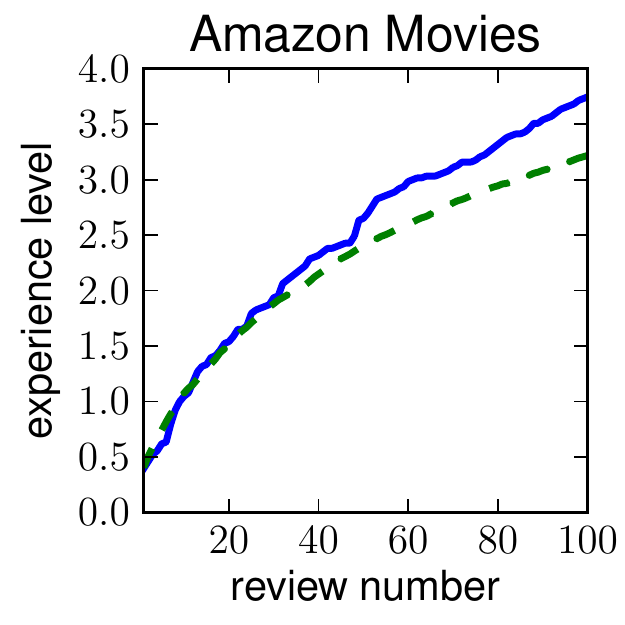}%
\includegraphics[scale=0.555]{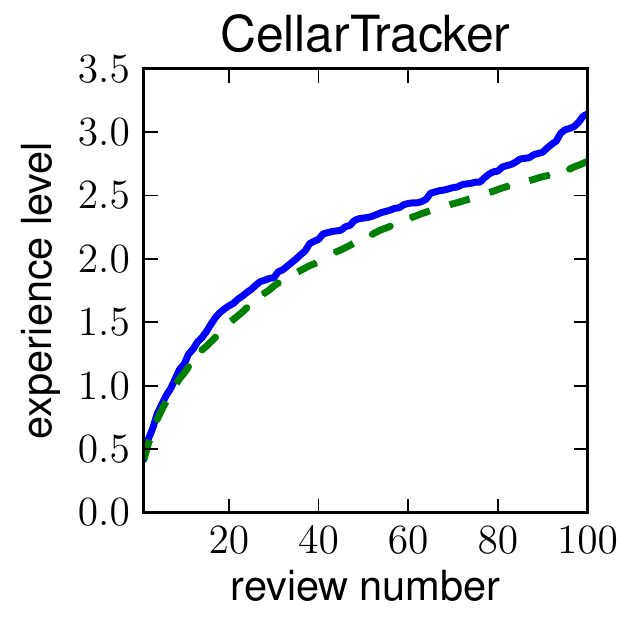}%

\vspace{-2mm}
\end{center}
\caption{Users whose experience progresses slowly are more likely to abandon the community. First 10 ratings of all users who have at least 10 ratings (top), and first 100 ratings of all users who have at least 100 ratings (bottom). \label{fig:expcurves}}
\vspace{3.3mm}
\end{figure*}

Next we consider how experience relates to user retention. We want to study how users who leave the community (defined as users who have not entered a review for a period of six months) differ from those who remain in the community. Figure \ref{fig:expcurves} visualizes the experience progression of these two groups. Here we consider the first 10 ratings for all users who have entered at least 10 ratings (Fig.~\ref{fig:expcurves}, top), and the first 100 ratings for all users who have entered at least 100 ratings (Fig.~\ref{fig:expcurves}, bottom); this scheme ensures that every datapoint is drawn from the same sample population.

We find that both classes of users \emph{enter} the community at roughly the same level (at the time of their first review, both groups have roughly the same experience on average). However, as the number of reviews increases, users who go on to leave the community have lower experience compared to those who stay. In other words, they gain experience more slowly. This discrepancy is apparent even after their first few reviews. This failure to become experienced may be a factor which causes users to abandon the site, and could be used as a feature in `churn prediction' problems \cite{dror12,churn}. We mention the parallel work of \cite{cdnm}, which also studies \emph{BeerAdvocate} and \emph{RateBeer} data: there, a user's failure to adopt the \emph{linguistic} norms of a community is considered as a factor that may influence whether they will abandon that community.

\subsection*{Acquired Tastes}
In Figure \ref{fig:genreVexp}, we hinted at the idea that our model could be used to detect \emph{acquired tastes}. More precisely, it can help us to identify products that are preferred by experts over beginners (and \emph{vice versa}).

To do so, we compare the difference in product bias terms between the most expert (experience level 5) and the least expert (experience level 1) users. That is, we compute for each item $i$ the quantity
\begin{equation*}
 d_i = \beta_i(5) - \beta_i(1).
\end{equation*}
Thus a positive value of $d_i$ indicates that a product is preferred by experts over beginners, while a negative value indicates that a product is preferred by beginners over experts.

\begin{figure*}[t]
\begin{center}
\includegraphics[width=0.2\textwidth]{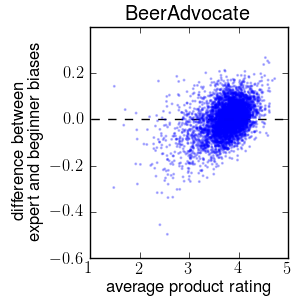}%
\includegraphics[width=0.2\textwidth]{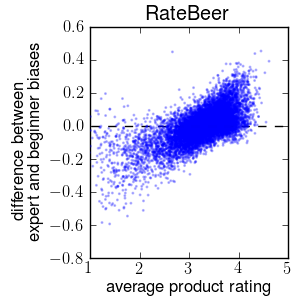}%
\includegraphics[width=0.2\textwidth]{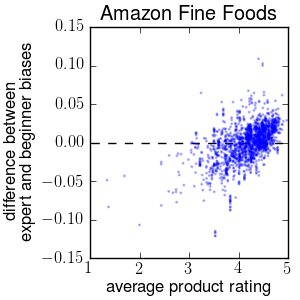}%
\includegraphics[width=0.2\textwidth]{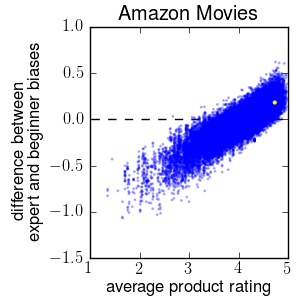}%
\includegraphics[width=0.2\textwidth]{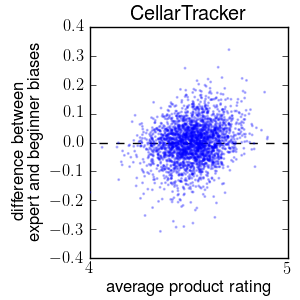}
\vspace{-4mm}
\end{center}
\caption{Average product ratings are correlated with the difference between expert and beginner biases. \emph{Seven Samurai} is marked in yellow. \label{fig:prodVexp}}
\end{figure*}

\xhdr{How do expert and beginner biases differ?}
In Figure \ref{fig:prodVexp} we compare the average rating of each product to $d_i$ (for products with at least 50 ratings). Our main finding in this figure is that there exists a positive relationship between products that are highly rated and products that are preferred by experts. In other words, products with high average ratings are rated \emph{more} highly by experts; products with low average ratings are rated more highly by beginners. Recall that in Figure \ref{fig:genreVexp} we examined the same relationship on \emph{RateBeer} data in more detail.

One explanation is that the `best' products tend to be ones that require expertise to enjoy, while novice users may be unable to appreciate them fully. This phenomenon is the most pronounced on our \emph{Movies} and \emph{RateBeer} data, and exists to a lesser extent on our \emph{BeerAdvocate} and \emph{Fine Foods} data; the phenomenon is absent altogether on our \emph{CellarTracker} data. Again, we should not conclude from this that movies require more `expertise' than wine, but rather that our \emph{Movies} data has a larger separation between beginners and experts.

\vspace{0.7mm}

Perhaps more surprising is the lack of products that appear in the top left or bottom right quadrants of Figure \ref{fig:prodVexp}, i.e., products with below average ratings, but positive values of $d_i$, or products with above average ratings but negative values of $d_i$. In other words, there are neither products that are disliked by beginners but liked by experts, nor are there products that are liked by beginners but disliked by experts.

\vspace{0.7mm}

It is worth trying to rule out other, more prosaic explanations for this phenomenon: for instance, it could be that beginners give mediocre reviews to all products, while experts have a larger range. We mention two negative results that discount such possibilities: firstly, we found no significant difference between the average ratings given by beginners or experts. Secondly, we did not observe any significant difference in the variance (that is, the variance across all of a user's reviews, not when reviewing the same product as in Figure \ref{fig:agreement}).

\xhdr{Which genres are preferred by experts or beginners?}
In Figure \ref{fig:genreVexp} we showed that there are entire \emph{genres} of products that tend to be preferred by experts or by beginners. Specifically, we showed that almost all strong ales have positive values of $d_i$ (preferred by experts), while almost all lagers have negative values of $d_i$ (preferred by beginners). Of course, it is not surprising (to a beer drinker) that experts dislike lagers while preferring India Pale Ales (IPAs), though it is more surprising that beginners also have the same polarity with respect to these products---the experts are simply more extreme in their opinions.

Table \ref{tab:genre} shows which genres have the lowest and highest values of $d_i$ on average, i.e., which products are most preferred by beginners and experts (respectively). We focus on \emph{BeerAdvocate}, \emph{RateBeer}, and \emph{CellarTracker}, which have the most meaningful genre information. The results are highly consistent across \emph{BeerAdvocate} and \emph{RateBeer}, in spite of the  differing product categorizations used by the two sites (Kvass is a form of low-alcohol beer, Kristallweizen is a form of wheat beer, IPA is a form of strong ale, and Gueuze is a type of lambic). Again, there is a clear relationship between products' overall popularity and the extent to which experts prefer them; non-alcoholic beer is naturally not highly rated on a beer rating website, while lambics and IPAs are more in favor.

\section{Related Work}
\label{sec:related}
Traditional recommender systems treat each user's review history as a series of unordered events, which are simply used to build a model for that user, such as a latent factor model \cite{Koren09}. In spite of the excellent performance of such models in practice, they naturally fail to account for the temporal dynamics involved in recommendation tasks.

Some early works that deal with temporal dynamics do so in terms of \emph{concept drift} \cite{kuncheva,Tsymbal04theproblem,widmer96}. Such models are able to account for short-term temporal effects (`noise'), and long-term changes in user behavior (`drift'), for example due to the presence of new products within a community.

Sophisticated models of such temporal dynamics proved critical in obtaining state-of-the-art performance on the \emph{Netflix} challenge \cite{netflix}, most famously in \cite{koren10}. As discussed in \cite{koren10}, few previous works had dealt with temporal dynamics, other than a few notable exceptions \cite{Bell07,Ding05,Sugiyama04}. Around the same time, `adaptive neighborhood' models were proposed \cite{Lathia09}, that address the problem of iteratively training recommender systems whose parameters ought to change over time.

Better performance may be obtained by modeling large-scale global changes at the level of entire communities \cite{xiong}, or by developing separate models for short term changes (e.g.~due to external events), and long-term trends \cite{xiang}.

Other works that study temporal dynamics at the level of products and communities include \cite{moe12}, where the authors studied how existing ratings within communities may influence new users, and how community dynamics evolve; and \cite{godes2012}, who studied how users are influenced by previous ratings of the same product.

Expertise has been studied in domains other than recommender systems, for example in the literature on education and psychology \cite{berliner88,romaine1984language}. One area where `expertise' has received significant attention is \emph{web search}. The role of expertise with respect to search behavior is a rich and historied topic, whose study predates the emergence of modern search engines \cite{ingrid94}. Of particular interest is \cite{white09}, since the authors study how users \emph{evolve} (with respect to the level of technical content in their queries) as they gain expertise. We also briefly mention the topic of \emph{expertise identification} \cite{alonso2008expertise,bian2009learning,jurczyk2007discovering,pal11b}. This line of work is orthogonal to ours, in that it deals with \emph{discovering experts}, rather than \emph{recommending products based on expertise}; however, such works offer valuable insights, in the sense that like our own work, they attempt to model the behavior of expert users.

\begin{table*}
\begin{center}
\ \\[3.7mm]
\small
\begin{tabular}{llrrlrrlrr}
                               & BeerAdvocate            & $\bar\beta_i(1)$ & $\bar\beta_i(5)$ & RateBeer & $\bar\beta_i(1)$ & $\bar\beta_i(5)$ & CellarTracker & $\bar\beta_i(1)$ & $\bar\beta_i(5)$\\
\hline\\[-2.7mm]
Preferred                    & Low Alcohol Beer        & -.423 & -.534 & Low Alcohol         & -.581 & -.724 & Barbera               & -.035  & -.119\\
by beginners:                & Kvass                   & -.316 & -.653 & Pale Lager          & -.519 & -.630 & Syrah (blend)         & -.110  & -.182\\
                             & Light Lager             & -.302 & -.487 & Premium Lager       & -.246 & -.290 & Cabernet-Syrah        & -.048  & -.105\\
                             & American Adjunct Lager  & -.237 & -.285 & American Dark Lager & -.199 & -.287 & Zinfandel             & -.066  & -.111\\
                             & European Strong Lager   & -.403 & -.428 & Strong Pale Lager   & -.068 & -.103 & S\'emillon            & -.092  & -.126\\
                             & European Pale Lager     & -.154 & -.216 & German Pilsener     & -.135 & -.202 & Syrah                 &  .019  &  .005\\
                             & Japanese Rice Lager     & -.144 & -.212 & Pilsener            & -.066 & -.095 & Port                  & -.028  & -.046\\
                             & American Pale Wheat Ale &  .033 & -.023 & Kristallweizen      &  .060 &  .016 & Grenache (blend)      &  .011  & -.005\\
                             & American Blonde Ale     & -.047 & -.080 & Fruit Beer          & -.105 & -.137 & Sauvignon Blanc       & -.003  & -.020\\
                             & English Dark Mild Ale   & -.025 & -.072 & Malt Liquor         & -.557 & -.675 & Gr\:uner Veltliner    &  .024  &  .008\\
\hline\\[-2.7mm]
Preferred                    & Baltic Porter           &  .091 &  .128 & Strong Porter       &  .205 &  .243 & Melon de Bourgogne     &  .261 &  .412\\
by experts:                  & English Barleywine      &  .007 &  .055 & Barley Wine         &  .216 &  .248 & Champagne              &  .080 &  .193\\
                             & American Wild Ale       &  .150 &  .196 & Wild Ale            &  .197 &  .261 & Cabernet-Syrah (blend) &  .083 &  .173\\
                             & English Pale Mild Ale   & -.011 &  .023 & American Strong Ale &  .203 &  .227 & Petit Verdot           &  .054 &  .143\\
                             & Flanders Red Ale        &  .132 &  .183 & Double IPA          &  .260 &  .294 & Pinot Gris             &  .118 &  .206\\
                             & Flanders Oud Bruin      &  .018 &  .073 & Black IPA           &  .152 &  .185 & Pinotage               &  .064 &  .117\\
                             & Unblended Lambic        & -.019 &  .028 & Unblended Lambic    &  .135 &  .240 & Grenache               &  .009 &  .048\\
                             & Gueuze                  &  .160 &  .223 & Saison              &  .142 &  .176 & Grenache Blanc         &  .038 &  .089\\
                             & Chile Beer              & -.254 & -.223 & Imperial Stout      &  .308 &  .338 & Dolcetto               &  .063 &  .105\\
                             & Rauchbier               & -.143 & -.095 & Quadrupel           &  .354 &  .367 & Mourvedre Blend        & -.031 &  .007\\
\end{tabular}
\normalsize
\vspace{-3mm}
\end{center}
\caption{Acquired tastes: products preferred by beginners and products preferred experts. Average beginner and expert item biases are shown.\label{tab:genre}}
\end{table*}

\section{Discussion and Future Work}
\label{sec:discussion}

An interesting finding of our work is that beginners and experts have the same \emph{polarity} in their opinions, but that experts give more `extreme' ratings: they rate the top products more highly, and the bottom products more harshly. Thus naively, we might conclude that we should simply recommend both groups of users the same products: nobody likes adjunct lagers, so what does it matter if beginners dislike them \emph{less}? The counter to this argument is that in order to \emph{fully} appreciate a product (by giving it the highest rating), a user must first become an expert. Thus perhaps we should focus on \emph{making a user an expert}, rather than simply recommending what they will like \emph{today}.

This viewpoint motivates several novel questions. Can we determine, based only on which products a user reviews, whether they will become an expert? Can we recommend not just products, but \emph{sequences} of products, that will \emph{help} them to become an expert, or maximize their total enjoyment?

Another avenue of research is to study \emph{linguistic} differences between experts and non-experts. `Expertise' has been studied from the perspective of linguistic development \cite{romaine1984language}, for example in the context of second-language acquisition \cite{cumming2006writing,thorne09}. Since our rating data comes from \emph{review} corpora, we can use it to study how users' \emph{rating} expertise relates to their \emph{reviewing} expertise. Do experts write longer reviews or use fewer personal pronouns? Are their reviews considered more helpful by others in the community \cite{nguyen2011language}? Do experts use specialized vocabulary (e.g.~beer-specific language such as `lacing' and `retention'), and do they better conform to the linguistic norms of the community \cite{cdnm}?

Although we argued that gaining expertise is a form of \emph{personal} development that takes place no matter when a user arrives in a community, it is not the \emph{only} such form of development. When a user arrives in a community, they must adopt its norms with respect to rating behavior \cite{moe12}, for example they must learn that while wines are rated on a scale of 1-100, ratings below 85 are seldom used. Like expertise, this form of development is not tied to the user's changing preferences, nor to the changing tastes of the community.
Would \emph{BeerAdvocate} `experts' be considered experts if their ratings were entered on \emph{RateBeer}?
Would expert movie reviewers on \emph{Amazon} also be `experts' gourmet food reviewers? The answers to such questions could help us to determine which forms of personal development are more salient in online communities.

Finally, we believe that our models of expertise may be used to facilitate expert discovery. We identified experts in review systems primarily for the sake of predicting users' ratings, but discovering experts may be useful in its own right. This topic is known as \emph{expert recommendation} \cite{alonso2008expertise,bian2009learning,craswell2001p,jurczyk2007discovering,karimzadehgan2009enhancing}, and has applications to product ranking and review summarization, among others. For instance, a user may wish to read reviews written by the most expert members of a community, or by members most similar in expertise to themselves.

\section{Conclusion}

Users' tastes and preferences change and evolve over time. Shifting trends in the community, the arrival of new products, and even changes in users' social networks may influence their rating behavior. At the same time, users' tastes may change simply through the act of consuming additional products, as they gain knowledge and experience. Existing models consider temporal effects at the level of products and communities, but neglect the \emph{personal development} of users: users who rate products at the same time may have less in common than users who rate products at \emph{different} times, but who are at the same stage in their personal evolution. We developed models for such notions of user evolution, in order to assess which best captures the dynamics present in product rating data. We found that modeling users' personal evolution, or `experience', not only helps us to discover `acquired tastes' in product rating systems, but more importantly, it allows us to discover \emph{when} users acquire them.

\subsection*{Acknowledgements} Thanks to Cristian Danescu-Niculescu-Mizil for help obtaining the data, and to Seth Myers and Dafna Shahaf for proofreading. 
This research has been supported in part by NSF
IIS-1016909,
CNS-1010921,
CAREER IIS-1149837,
IIS-1159679,
ARO MURI,
Docomo,
Boeing,
Allyes,
Volkswagen,
Intel,
Okawa Foundation,
Alfred P. Sloan Fellowship and
the Microsoft Faculty Fellowship.


\balancecolumns
\vfill\eject

\end{document}